\documentclass[thmrestate,cleveref, autoref, thm-restate]{article}

\usepackage{thm-restate}
\usepackage{xspace} 
\usepackage{textcomp}
\usepackage{balance}
\usepackage{cite}
\usepackage{balance}
\usepackage{url}
\usepackage{amsthm}
\usepackage{mathtools}
\usepackage{fullpage}

\newtheorem{observation}{Observation}

\newcommand{\defn}[1]       {{\textit{\textbf{\boldmath #1}}}}

\renewcommand{\epsilon}{\varepsilon}

\newcommand{\consta}{3}

\newcommand{\jellotree}{J$\epsilon$llo Tree\xspace}

\newcommand{\bpt}{buffered propagation tree\xspace}

\newcommand{\bpts}{buffered propagation trees\xspace}

\newcommand{\BPTS}{Buffered Propagation Trees\xspace}

\newcommand{\seclabel}[1]    {\label{sec:#1}}

\renewcommand{\eqref}[1]          {Eq.~\ref{#1}}

\title{What Does Dynamic Optimality Mean in External Memory?\thanks{Portions of this work were completed at the Second Hawaii Workshop on Parallel
Algorithms and Data Structures. The authors would like to thank Nodari Sitchinava for organizing
the workshop.}}

\author{Michael A. Bender\thanks{Supported in part by NSF grants 
CCF-2118832, 
CCF-2106827, 
CCF-1725543, 
CSR-1763680, 
CCF-1716252, 
and
CNS-1938709.   
}\\Stony Brook University, Stony Brook, NY, USA\\\texttt{bender@cs.stonybrook.edu} \and
Mart\'{\i}n Farach-Colton\thanks{Supported in part by NSF grants  CSR-1938180, 
CCF-2106999, and CCF-2118620.
}\\Rutgers University, New Brunswick, NJ, USA\\\texttt{martin@farach-colton.com} \and William Kuszmaul\thanks{Funded by a Hertz Fellowship and an NSF GRFP Fellowship. Research was also partially sponsored by the United States Air Force Research Laboratory and the United States Air Force Artificial Intelligence Accelerator and was accomplished under Cooperative Agreement Number FA8750-19-2-1000. The views and conclusions contained in this document are those of the authors and should not be interpreted as representing the official policies, either expressed or implied, of the United States Air Force or the U.S. Government. The U.S. Government is authorized to reproduce and distribute reprints for Government purposes notwithstanding any copyright notation herein.}\\MIT, Cambridge, MA, USA\\\texttt{kuszmaul@mit.edu} }

\begin{document}

\maketitle
\begin{abstract}

  A data structure $A$ is said to be dynamically optimal over a class
  of data structures $\mathcal{C}$ if $A$ is constant-competitive
  with every data structure $C \in \mathcal{C}$. Much of the research
  on binary search trees in the past forty years has focused on
  studying dynamic optimality over the class of binary search trees
  that are modified via rotations (and indeed, the question of whether splay trees are dynamically optimal has gained 
  notoriety as the so-called dynamic-optimality conjecture).
  Recently, researchers have extended
  this to consider dynamic optimality over certain classes of
  external-memory search trees. In particular, Demaine, Iacono,
  Koumoutsos, and Langerman propose a class of external-memory trees
  that support a notion of tree rotations, and then give an elegant
  data structure, called the Belga B-tree, that is within an
  $O(\log \log N)$-factor of being dynamically optimal over this
  class.

In this paper, we revisit the question of how dynamic optimality should be defined in external memory. A defining characteristic of external-memory data structures is that there is a stark asymmetry between queries and inserts/updates/deletes: by making the former slightly asymptotically slower, one can make the latter significantly asymptotically faster (even allowing for operations with sub-constant amortized I/Os). This asymmetry makes it so that rotation-based search trees are not optimal (or even close to optimal) in insert/update/delete-heavy external-memory workloads. To study dynamic optimality for such workloads, one must consider a different class of data structures.

The natural class of data structures to consider are what we call buffered-propagation trees. Such trees can adapt dynamically to the locality properties of an input sequence in order to optimize the interactions between different inserts/updates/deletes and queries. We also present a new form of beyond-worst-case analysis that allows for us to formally study a continuum between static and dynamic optimality. Finally, we give a novel data structure, called the \jellotree, 
that is statically optimal and that achieves dynamic optimality for a large natural class of inputs defined by our beyond-worst-case analysis.  
\end{abstract}

\section{Introduction} \seclabel{intro}

\paragraph*{Static and dynamic optimality in internal memory} Since the early 1960s, many balanced binary trees have been developed with worst-case $O(\log N)$ time per operation~\cite{AdelsonLa62, Bayer72, CormenLeRi01}, where $N$ is the number of elements in the tree.  In such trees, the cost of any particular operation can be much smaller, even $O(1)$, if the element being queried is stored near the root of the tree. This means that a search tree can potentially achieve $o(\log N)$ time per operation on workloads that exhibit locality properties.

Since the 1980s, considerable effort has been devoted to designing \defn{distribution-sensitive binary search trees} that perform workload-specific optimizations.  Broadly speaking, there are two approaches to analyzing distribution-sensitive search trees. The first approach is to bound the performance based on some property of the input sequence~\cite{Tarjan85, SleatorTa85b, Iacono01, ColeMiScSi00, ColeMiScSi00b, BoseDoIaLa14, BuadoiuCoDeIa07, HowatIaMo13, ChalermsookGKMS18}, e.g., the sequential access bound \cite{Tarjan85}, the working set bound \cite{SleatorTa85b, Iacono01}, the weighted dynamic finger bound~\cite{ColeMiScSi00, ColeMiScSi00b, BoseDoIaLa14}, and the unified bound \cite{BuadoiuCoDeIa07, Iacono01}. 
The second approach is competitive analysis, where one must select a class of data structures $\mathcal{C}$, such as static binary trees or binary trees that are modified via rotations, and then design a single data structure $A$ (not necessarily from $\mathcal{C}$) that is competitive with any data structure in $\mathcal{C}$.  If the members of $\mathcal{C}$ are static, a $O(1)$-competitive algorithm\footnote{Here, we can see an example where it is especially important that $A$ not have to be a member of $\mathcal{C}$. In particular, if we wish to construct an $A$ that is $O(1)$-competitive with any (omnisciently constructed) static $C$, then we must allow for $A$ to adapt dynamically over time.} is said to be \defn{statically optimal against $\mathcal{C}$} and if they are dynamic, a $O(1)$-competitive algorithm is said to be \defn{dynamically optimal against $\mathcal{C}$}.

Splay trees are famously statically optimal against the class of binary trees~\cite{SleatorTa85b}. On the other hand, whether dynamic optimality can be achieved against the class of binary-trees-with-rotations remains one of most elusive problems in the field of data structures (see \cite{Iacono13} for a survey). The Tango Tree~\cite{DemaineHaIaPa07} is known to be within a factor of $O(\log \log N)$ of dynamic optimality. The splay tree~\cite{SleatorTa83} is widely believed to achieve dynamic optimality, but it remains open whether the structure is even $o(\log N)$-competitive.

Research on dynamic optimality against internal-memory search trees has historically considered sequences of operations consisting exclusively of queries. We emphasize that this is \emph{not} a limitation of past work---indeed,  it turns out that queries and inserts/updates/deletes are sufficiently similar to one another that the queries-only assumption is without loss of generality.  As we will see later, this equivalence does not hold in external memory.

\paragraph*{Static and dynamic optimality in external memory} Search trees are every bit as ubiquitous in external memory as they are in internal memory---e.g., they   are used prominently in file systems~\cite{FS1,FS2}, databases~\cite{BayerMc72, Comer79}, and key-value stores~\cite{ConwayFa18,rocks,pebbles}.  The principle difference with internal memory is that disks are accessed in blocks of some (typically large) size $B$.\footnote{As a convention, $B$ is measured in terms of the number of machine words that fit in a block.} External-memory search trees are analyzed in the \defn{Disk-Access Model}~\cite{AggarwalVi88}, where the goal is to minimize the number of block accesses (also known as \defn{I/Os}).

B-trees~\cite{BayerMc72, Comer79} are balanced search trees optimized for external memory, which means they have fanout $\Theta(B)$, and hence a worst case of $O(\log_B N)$ I/Os per operation.  Again, the cost of any particular operation can be much smaller if the element being queried is stored near the root.  This raises a natural question: what can be said about static and dynamic optimality for external-memory search trees? We remark that most of the work on this question has focused on dynamic optimality and treated static optimality implicitly.
  
The notion of dynamic optimality in external-memory search trees is less well understood than in internal memory. Part of the reason for this is the difficulty of identifying the class of data structures over which dynamic optimality should be defined, and in particular identifying the mechanism by which elements can move vertically in the tree.
   
Early work focused on a skip-list-like mechanism where keys can move towards the root in the tree by becoming a pivot and splitting some node in two~\cite{Sherk95, BoseDL08}.  We call this class of data structures \defn{merge-split trees}.  Bose et al. \cite{BoseDL08} gave a data structure that achieves dynamic optimality in this model and showed that the performance of their data structure is determined by the working-set bound \cite{SleatorTa85b, Iacono01}.

Merge-split trees are limited in their ability to exploit the locality of a workload. For example, on the sequential workload in which $1, 2, \ldots, N$ are accessed round robin, merge-split trees incur amortized cost $\Omega(\log_B N)$ per operation. In contrast, even in-memory binary trees implemented using rotations can achieve the \defn{sequential access bound} \cite{Tarjan85} on the same workload, that is, an amortized $O(1)$ I/Os. The main difference between merge-split trees and binary trees with rotations is that merge-split trees move individual elements up and down the tree, but they do not move entire subtrees together (as is the case for rotations).

Recently, Demaine et al.~\cite{DemaineIKL19} introduced a different class of dynamic trees that support ``rotations'' similar to those in in-memory binary search trees. The authors study dynamic optimality over this class of data structures and introduce the Belga B-tree, which they prove is $O(\log \log N)$-competitive against any rotation-based search tree.

The work on external-memory dynamic optimality so far has focused on exploiting the underlying locality properties of the workload in order to optimize queries. The tradeoff being explored is the decision of which keys are stored near the root of the tree and which keys are stored further down.

\paragraph*{This paper: optimizing the asymmetry of external-memory operations}  One of the remarkable (and perhaps unexpected) differences between search trees in internal and external memory, however, is that in external memory, inserts/updates/deletes can be implemented to have amortized performance asymptotically faster than that of queries.  While the worst-case cost of queries is logarithmic, inserts/updates/deletes can take an amortized subconstant number of I/Os~\cite{BrodalFa03c, BenderFaJaKu15, BenderFaFi07, ONeilChGa96, BuchsbaumGoVeWe00, bender2020flushing}. 

An important consequence of this asymmetry is that there are many input sequences for which the positioning of different keys in the tree is \emph{not} the dominant factor controlling performance. To study dynamic optimality for such sequences, we must consider a class of algorithms that can optimize the cost of queries vs insertion/deletions/updates.  

The key technique for such optimizations is \defn{buffered propagation} in which one propagates insert/update/delete operations down the tree in buffered batches. This allows for a single I/O to make progress on many insert/update/delete operations simultaneously, so that the amortized cost of such operations is small. We emphasize that, on insert/update/delete-heavy workloads, even standard trees that \emph{non-adaptively} use buffered propagation, such as the $B^{\epsilon}$-trees~\cite{BrodalFa03c, BenderFaJaKu15, arge1995buffer, bender2020flushing}, can be asymptotically faster than the best possible adaptive rotation-based search trees. 

Buffered propagation comes with a tradeoff curve: we can make inserts/updates/deletes faster (up to a factor of $O(B)$) at the cost of making queries slower (up to a factor of $O(\log B)$).  This means that there is an opportunity to adapt dynamically to a sequence of operations, both by adjusting the amount of buffered propagation over time, and by using \emph{different amounts of buffered propagation in different parts of the tree}. There is also an opportunity to adapt the choice of pivots used by each internal node in the tree in order to strategically split collections of operations in a way that sends queries in one direction and inserts/updates/deletes in another.  

All of these tradeoffs can be formally captured with a class of data structures that we call \defn{buffered-propagation trees}---the problem of optimizing the tradeoffs between queries and non-queries in an external-memory search tree corresponds to the problem of achieving dynamic optimality against buffered-propagation trees.

We remark that even \emph{static optimality} against \bpts is an interesting question. That is, given a workload of query operations and update operations (which change values associated with keys, so the set of keys does not change over time), can one construct a data structure that is competitive with the optimal statically-structured \bpt? Even though the \bpt has a static structure, it can still strategically select the pivots and the amount of buffered propagation at each internal node in order to optimize for spatial locality between operations. Even if we are given the operations up front (in an offline manner), it is not immediately clear how one should go about constructing the optimal static \bpt. 

\paragraph*{A continuum between static and dynamic optimality}
Achieving full dynamic optimality against any sophisticated class of search trees (whether it be internal-memory rotation-based trees or external-memory buffered-propagation trees) is a difficult problem to get traction on: even small changes to a tree can have significant impact on asymptotic performance, so an omniscient adversary can potentially perform rapid modifications to the data structure in order to adapt to the workload at a very fine-grained level. In addition to considering the question of how to model dynamic optimality in external memory, a second contribution of this paper is to revisit the question of how we should perform beyond-worst-case analysis within that model, in order to characterize how ``close'' a given data structure is to achieving dynamic optimality.

One insight is that, in practice, it is natural to expect that the properties of an input sequence may evolve slowly over time, meaning that the (offline) optimal dynamic \bpt will also evolve slowly. We capture this property formally by declaring a sequence of operations to be \defn{$K$-smooth} if there exists an optimal dynamic \bpt $T$ for the sequence such that only a $1 / K$-fraction of $T$'s I/Os are spent restructuring the tree.

We propose a natural form of beyond-worst-case analysis: rather than trying to achieve full dynamic optimality, can we achieve dynamic optimality for the class of $K$-smooth inputs (and for some reasonably small $K$)? Instead of thinking of this as a restriction on input sequences, one can also think of it as a type of resource augmentation. Can we design a data structure that is $O(1)$-competitive with any \defn{$K$-speed-limited \bpt}, that is any \bpt that is limited to spend at most a $1/K$-fraction of its I/Os on modifying the tree? Note that any data structure that is $O(1)$-competitive against $K$-speed-limited \bpts is guaranteed to be $O(1)$-competitive on all $K$-smooth input sequences---therefore the problem of achieving competitive guarantees against $K$-speed-limited adversaries subsumes the problem of achieving optimality on $K$-smooth inputs.

The study of $K$-speed-limited adversaries offers an intriguing continuum between static and dynamic optimality. Optimality against $\infty$-speed-limited adversaries is equivalent to static optimality, and optimality against $1$-speed-limited adversaries is equivalent to full dynamic optimality. The smaller a $K$ that we can achieve optimality against, the closer we are to achieving true dynamic optimality. We remark that this same continuum would also be interesting to study for internal-memory search trees, and we leave this direction of work as an open problem.

\paragraph*{Achieving dynamic optimality against a speed-limited adversary} 
The third contribution of the paper is a new data structure that we call the \jellotree. The \jellotree is statically optimal, meaning that on any workload of updates and queries, the tree is constant competitive with any static \bpt. The \jellotree is also dynamically optimal against any sufficiently speed-limited \bpt.  

Our main theorem is that, for any $\delta \in (0, 1) \cap \Omega(\log \log N / \log B)$, we can build a \jellotree that is $O(1/\delta)$-competitive with any $B^{\consta \delta}$-speed-limited \bpt. The construction and analysis of the \jellotree is the main technical result of the paper.

In addition to requiring that the adversary is speed limited, our competitive analysis assumes a $B^{\consta \delta}$-factor of resource augmentation on cache size, meaning that we compete an adversary whose cache is a $B^{\consta \delta}$-factor smaller than ours. We also present a version of the analysis that incurs a small additive overhead in exchange for eliminating the cache-size resource augmentation.

\paragraph*{Paper outline} In Section \ref{sec:opt}, we formally define \bpts and speed-limited adversaries, and we state our main results. Then in Section \ref{sec:technical} we give a sketch of the \jellotree's design and analysis. The full design and analysis appear in the extended version of the paper \cite{arxiv}.

\section{Defining the Class of Speed-Limited \BPTS}\label{sec:opt}
In this section, we formally define
\bpts (and speed-limited \bpts). We then define the adversary Speed-Limited
OPT (or OPT for short) against which we will analyze the
\jellotree. And finally we use these definitions to formally state the main result of the paper.

Both \bpts and the \jellotree live in
the Disk-Access Model \cite{AggarwalVi88}. In particular, the computer has a cache of size $M$ machine words,
and has access to an (unbounded-size) external memory consisting of blocks of some size $B$ machine words.
An algorithm can read/write a block from external memory to cache at the cost of one \defn{block access} 
(or \defn{I/O}), and time is measured as the total number of I/Os incurred by the algorithm.

\subsection{An introduction to buffered propagation}

In external-memory data structures, there is an asymmetry that allows for inserts/updates/deletes to be implemented asymptotically faster than query operations. The fundamental technique for achieving these speedups is \defn{buffered propagation}, in which one propagates insert/update/delete operations down the tree in buffered batches. For each node $x$ in the tree, if $x$ has $f$ children $c_1, \ldots, c_f$, then $x$ maintains a buffer of size $B / f$ for each of those children. Each buffer collects insert/update/delete messages destined for that child $c_i$. Messages are flushed down from $x$ to the children $c_1, \ldots, c_f$ in collections of size $B / f$ (i.e., whenever a buffer for one of the children overflows). The $O(1)$ I/Os that are used to perform a buffer flush are shared across $\Theta(B / f)$ insert/update/delete operations. By giving a node $x$ a smaller fanout, one can decrease the amortized cost of a buffer flush, making inserts/updates/deletes faster. On the other hand, smaller fanouts also make the height of the tree larger, which makes queries slower.

A classic example of buffered propagation is the $B^{\epsilon}$-tree \cite{BuchsbaumGoVeWe00, BenderFaJaKu15, BrodalFa03c, bender2020flushing}, which has found applications in databases~\cite{BenderFaJaKu15, Tokutek09, TokuDB11} and file systems~\cite{FS1, FS2, FS3, FS4, FS5, FS6}.  In a $B^{\epsilon}$-tree, all nodes have the same fixed fanout $f$ (typically, one sets $f = B^{\epsilon}$ for some constant $\epsilon$). Queries cost $O(\log_f n)$ and insert/update/delete operations have amortized cost $O(\frac{f}{B} \log_f N)$, so that, e.g., an insert/update/delete-heavy workload can be performed asymptotically faster than in standard $B$-trees if $f$ is selected to be small. An interesting feature of this tradeoff curve is that, if $B \gg f \log_f N$, then insert/update/delete operations can even take \emph{sub-constant} amortized time---the same guarantee is not possible for queries.

The fanout $f$ used within a $B^{\epsilon}$-tree can be tuned to the sequence of operations. One must be careful not to select the wrong fanout for the workload, however. For example, the $B^{1/2}$-tree, where $f = \sqrt{B}$, achieves an  insert/update/delete performance $O(\frac{\log_B n}{\sqrt{B}})$ while achieving an optimal query performance of $O(\log_B n)$. But, if a workload consists exclusively of inserts/updates/deletes then the $B^{1/2}$-tree will perform a factor of $\sqrt{B}$ away from optimal.
   
Although we typically think of $B^{\epsilon}$-trees as having fanout $f$ that is uniform across all nodes (and unchanging), the $B^{\epsilon}$-tree generalizes to a class of data structures where different nodes can have different fanouts. In this paper we define a broad class of data structures that we call \defn{\bpts}, and which can be viewed as weight-balanced $B^\epsilon$-trees with non-uniform fanout.

Non-uniform fanouts are especially natural if some parts of the key space are insert/update/delete heavy and other parts of the key space are query heavy.  A \bpt can pick a large $f$ for nodes that see mostly queries and a small $f$ for nodes that see mostly inserts/updates/deletes.  This means that pivots can be strategically selected in order to split collections of operations in a way that sends queries in one direction and inserts/updates/deletes in another.\footnote{As we will see later in this paper, the careful selection of pivots can have substantial asymptotic impact on the performance of the tree, even when fanouts are selected optimally.}  By contrast, since $f$ is uniform in a $B^\epsilon$-tree, the performance of the tree is fairly insensitive to pivot choice.  By choosing the right pivots and local $f$, one can potentially exploit the underlying spatial locality of the workload to achieve asymptotic improvements over any uniform-fanout $B^{\epsilon}$-tree.  In the dynamic case, where a \bpt is permitted to change its structure over time, it can also adapt to the temporal locality of the sequence of operations being performed.

\subsection{Formally defining \bpts}

We now formally define the class of \bpts. To simplify discussion, we restrict ourselves to queries/inserts/updates---discussion of deletes can be found in the extended version of the paper \cite{arxiv}.

For each node $x$ in a \bpt, let $\mathcal{K}(x)$ be the keys stored in the subtree rooted at $x$. Let $d(x)$ be the number of children that $x$ has and call them $c_1, \ldots, c_{d(x)}$.  Then $x$ selects some subset  $p_1 < \cdots < p_{d(x) - 1} \in \mathcal{K}(v)$ to act as \defn{pivots}. The children $c_1, \ldots, c_{d(x)}$ of $x$ then have key sets $\mathcal{K}(c_1) = \mathcal{K}(x) \cap (-\infty, p_1],  \mathcal{K}(c_2) = \mathcal{K}(x) \cap (p_1, p_2], 
\ldots, \mathcal{K}(c_j) = \mathcal{K}(x) \cap  (p_{j - 1}, \infty]$. The result is that each node $x$ is associated with some interval of keys, called $x$'s \defn{key range}, such that any operation on that key range is routed through $x$.

The \defn{size} of a node is the number of keys in the node's  key range in
the tree.\footnote{This can differ from $|\mathcal{K}(x)|$ for node $x$ because insertions into $x$'s key range can reside in a buffer above $x$.} 
Every node in the tree has a \defn{target size} (for leaves
the target size is $B / 2$), dictating what size the node is supposed to be: as a rule, at any given moment, if a node has target size $s$, then its true size must be in the range $[s, O(s)]$. All the children of a node must have the same target sizes as one another, and we refer to the target size of the children of a node as its \defn{target child size}. Without loss of generality, the target child size of a node is smaller than the target size.

\paragraph*{Flushing messages between nodes in a \bpt}
The \defn{target fanout} of an internal node is defined to be the
target size divided by the target child size. If a node has target
fanout $f$, then the node maintains a buffer of size $B / f$ for each
of its children. The buffer for each child $c$ stores insert/update messages
for that child---these messages keep track of insert/update
operations that need to be performed on keys in $c$'s key range. 

To understand how insert/update messages work, it is helpful to think about
the progression of a given message down the tree.
Any given insert/update operation on some key $k$ inserts a message into a
buffer at the root. Over time, the message then travels to the leaf
whose key range contains $k$, at which point the insert/update
operation is finally applied. Whenever a buffer for some child $y$
overflows in a node $x$, that buffer is flushed to the child $y$; and the 
messages in the buffer are distributed appropriately across $y$'s buffers;
this may then cause buffers in $y$ to overflow, etc..

In order to perform a query on a key $k$, one traverses the
root-to-leaf path to the leaf $\ell$ that contains $k$ in its key range. 
By examining the messages in the buffers of the nodes in the
root-to-leaf path, as well as the contents of leaf $\ell$, the tree
can answer the query on key $k$.

\paragraph*{Modifying a \bpt and defining speed-limitation}
A \bpt can dynamically change the fanouts and
pivot-choices within nodes in order to adapt to the sequence of
operations being performed. 

The most basic operation that a \bpt can perform
is to split a node $x$ into two nodes $x_1, x_2$ whose
target-sizes/target-child-sizes are the same as $x$'s were. This is
known as a \defn{balanced split}. Balanced splits allow for the tree
to perform weight-balancing, and we will treat balanced splits as
being free (for our adversary), meaning they do not cost any I/Os, 
even if the tree is $K$-speed-limited for some $K$.

The other way that a \bpt can modify itself is
through \defn{batch rebuild}, in which some collection of nodes
in the tree are replaced with new nodes (using possibly different
pivots and fanouts than before).

In more detail, when performing a batch rebuild, we can take any set of nodes
$X = \{x_1, \ldots, x_m\}$, and and replace them with a different
set of nodes $Y = \{y_1, \ldots, y_{m'}\}$ arbitrarily, with the restriction that 
after the replacement, the tree should still be valid (i.e., 
each node meets its target size requirement, each child has target size equal to the
parent's child target size, and pairs of consecutive key ranges are
separated by a valid pivot). Note that after a batch rebuild, buffers
may be significantly overflowed in some nodes, in which case the tree
must perform a series of buffer flushes to fix this. If a buffer is
overflowed by a factor of $k$, then flushing that buffer takes $\Theta(k)$ I/Os.\footnote{We can also think of the flush as being partitioned into $\lfloor k \rfloor$ distinct flushes, each of which flushes $B / f$ items.} 
  
A $K$-speed-limited \bpt is limited as follows: the tree is only permitted to devote a
$\frac{1}{K}$ fraction of its I/Os to batch rebuilds. Another way to think about this is that
I/Os spent on batch rebuilds are a factor-of-$K$ more expensive than other I/Os. So a
batch rebuild of a set of nodes $X$ into a new set of nodes $Y$ costs $(|X| + |Y|)K$ I/Os.

\paragraph*{Defining $K$-smooth inputs}
A \defn{$K$-smooth input} is any sequence of operations with the following property: The optimal \bpt
cost $C$ for those operations is within a constant factor of the optimal $K$-speed-limited \bpt cost
$C'$ for those operations. Intuitively, this means that there is an optimal (or at least near-optimal) 
\bpt that, during the sequence of operations, spends only a $O(1 / K)$-fraction of its I/Os on optimizing
the structure of the tree for the sequence.

Note that we are intentionally generous in what we consider to be ``optimizing the structure of the tree''.
Balanced splits are \emph{not} counted against the adversary, are not affected by $K$-speed-limitation, and
thus do not factor into $K$-smoothness. This is important because on an insertion-heavy workload, a tree may 
be \emph{forced} to perform a large number of balanced splits, even if the tree is not changing its fanouts/pivots
in any interesting way. Thus, insertion-heavy workloads would penalize the adversary unfairly for I/Os
that the adversary has no choice but to spend. 

To achieve dynamic optimality for $K$-smooth inputs, it suffices to achieve dynamic optimality (for all inputs) against
$K$-speed-limited \bpts:
\begin{observation}
If a data structure $T$ is $c$-competitive against dynamic $K$-speed-limited \bpts, then $T$ is $O(c)$-competitive 
on $K$-smooth input sequences against  dynamic \bpts.
\end{observation}
Throughout the body of the paper, we shall focus on the problem of achieving dynamic optimality against $K$-speed-limited \bpts,
since this problem is strictly more general than the problem of considering $K$-smooth inputs.

\paragraph*{Defining Speed-Limited OPT} In this paper, we will consider
the class of $K$-speed-limited \bpts, where
$K = B^{\consta \delta}$ for some parameter $\delta$. Given a sequence
of operations $S$, we define speed-limited OPT (or OPT for short) to
be the $K$-limited \bpt that achieves the minimum
total I/O cost on that sequence of operations. We will
design a data structure, the \jellotree, that is competitive with OPT.

\paragraph*{Caching in OPT and in the \jellotree}
OPT is assumed to a have a cache that stores the top of OPT's tree. As
per the Disk-Access Model \cite{AggarwalVi88}, any accesses to nodes
that are cached are free, in the sense that they do not incur I/Os. We
assume that the cache for OPT stores any node $x$ whose target size is
above $N / C$, for some caching parameter $C$. (Note that this means,
w.l.o.g., that OPT may as well make such nodes be fully insert/update-optimized
with $\Theta(1)$ fanouts---and furthermore, w.l.o.g., OPT does not
perform batch rebuilds on cached nodes.) We will further assume $N/ C = N^{1- \Omega(1)}$ (meaning that if OPT
were to have uniform fanout, then at most a constant fraction of OPT's
tree levels would be cached).

In the same way that we assume a factor of $B^{\consta \delta}$
resource augmentation in terms of speed-limitation, the \jellotree
will is also given a factor of $B^{\consta \delta}$ cache-size
resource augmentation against OPT. Namely, we will assume that the
\jellotree caches any node $x$ whose size is above
$\frac{N}{C B^{\consta \delta}}$. If both data structures were
$B^{\delta}$-trees, this would correspond with caching $O(1)$ more
layers than OPT caches. 

\subsection{Results}

Our main theorem is the following:

  \begin{restatable}{theorem}{thmmain}
    Suppose that $B \ge \Omega(\log N)$ and that $B$ is sufficiently
    large as a function of $1/\delta$. Let $\alpha$ be the total I/O
    cost incurred by the \jellotree, and let $\beta$ be the total cost
    incurred by the optimal $B^{\consta\delta}$-speed-limited \bpt OPT using a factor of $B^{\consta \delta}$ smaller cache than does the \jellotree. Then
    $\alpha \le O(\beta / \delta)$.
    \label{thm:main}
  \end{restatable}
  
We also present a version of the theorem that does not assume resource augmentation on cache size.  As long as $N \gg B$, then the cost of removing the resource augmentation is only a small additive I/O cost per operation. 
  
  \begin{restatable}{theorem}{thmmaintwo}
    Suppose that $B \ge \Omega(\log N)$ and that $B$ is sufficiently
    large as a function of $1/\delta$.  Let $\alpha$ be the total I/O
    cost incurred by the \jellotree, let $I$ be the total number of
    inserts/updates performed on the \jellotree, and let $R$ be the
    total number of queries performed on the \jellotree. Let $\beta$ be the total cost
    incurred by the optimal $B^{\consta\delta}$-speed-limited \bpt OPT using the same cache size as the \jellotree uses. Then
    $$\alpha \le O(\beta / \delta) + \min\{I/B^{\delta}, R \log B^{\delta}\}.$$
    \label{thm:main2}
  \end{restatable}
  
 In the extended version of the paper \cite{arxiv}, we also discuss how to incorporate deletes into both the definition of a speed-limited adversary and the design and analysis of the \jellotree.

\section{Technical Overview}\label{sec:technical}
Because both the \jellotree itself and its analysis are quite intricate, in this section we present a sketch of the main ideas in the data structure and our proofs. The full data structure and its analysis appear in the extended version of the paper \cite{arxiv}.

To simplify the presentation, we begin by considering optimality against a weakened version of OPT.  As subsections proceed, we remove restrictions on OPT and work our way towards achieving dynamic optimality the optimal $B^{\consta \delta}$-speed-limited \bpt.

We begin in Subsection~\ref{sec:epsilonconv} by considering an OPT that has 
uniform fanouts (i.e., OPT is a $B^{\epsilon}$-tree with optimal fanout). 
In Subsection~\ref{sec:fixed-pivot}, we consider an OPT that is allowed arbitrary fanouts, but
is restricted in its ability to select pivots. 
In Subsection~\ref{sec:technical-pivot-selection}, 
we examine the obstacles that arise OPT is permitted to select pivots arbitrarily. Finally, in Subsection \ref{sec:technical-supernode}, 
we consider the full version of OPT, in which OPT gets to select both pivots and fanouts freely. 

\subsection{A warmup: designing a \boldmath  fanout-convergent tree}\label{sec:epsilonconv}
Suppose we are given an initial set of $N$ records to be stored in a \bpt $T$ with $L = \Theta(N / B)$ leaves, and we are given a sequence of
operations $S = \langle s_1, s_2, \ldots, s_k \rangle$ of inserts/updates and queries.  Let $C_f(S)$ be the cost that the operations $S$ would incur
if $T$ were implemented as a $B^{\epsilon}$-tree with fanout $f$.
In this section, we present the \defn{fanout-convergent tree}, which 
is a data structure for implementing the operations $S$ so that the total 
cost is $O(\min_f C_f(S))$ (without knowing $S$ ahead of time).

\paragraph*{Problem: the cost of rebuilds}
A natural approach to achieving cost $O(\min_f C_f(S))$ would be to treat the selection of $f$ as a multi-armed
bandit problem \cite{AuerCeFrSc95}. The problem with this approach is
that, in order to offset the costs of rebuilding the tree in each
trial of the multi-armed bandit problem, each individual trial must be
very long. The result is that, in the time that it takes for the tree to
change size by a constant factor we would only be able to perform a small number of trials,
preventing the multi-armed bandit algorithm from
converging fast enough to be useful.

\paragraph*{Saving time by moving in only one direction}
In order to keep the total costs of tree rebuilds small, we only adjust the fanout $f$ in one direction. The tree
begins as fully query-optimized, i.e., with fanout $B$, and over time the
fanout decreases monotonically. A key insight is that, whenever a tree
with fanout $f$ is rebuilt as a new tree with smaller fanout, the
number of I/Os needed to perform this is only $O(L / f)$, since only
the \emph{internal} nodes of the tree need to be reconstructed. It
follows that if a tree begins with fanout $B$, and each successive
rebuild shrinks the fanout by a factor of at least two, then the total
cost of all of the rebuilds is a geometric series bounded by $O(L)$
I/Os. (Here we are treating the size of the tree as staying $O(N)$ at all times, 
but as we shall see momentarily, this assumption is without loss of generality.)

\emph{Any} \bpt must incur at least $\Omega((\log N) /B)\ge \Omega(1 / B)$ cost per insert/update and at least $\Omega(\log_B N) \ge \Omega(1)$ cost per query. Thus, whenever either (a) the total
number of inserts/updates surpasses $N$ or (b) the total number of
queries surpasses $L$, then the $O(L)$ cost of rebuilds has been amortized away.
Whenever either (a) or (b) occurs, we restart the entire procedure from scratch, returning to a fanout of $B$.

Since we restart our data structure each time that one of (a) or (b) occurs, we can assume without loss of generality that
the number of inserts/updates in $S$ is at most $N$, that the number of queries in $S$ is at most $L$, and that one of the two inequalities is strict (there are either exactly $N$ inserts/update or exactly $L$ queries). Given such an $S$, our challenge is to decrease the fanout over time in such a way that we achieve total cost $O(\min_f C_f(S))$.

\paragraph*{Selecting query-biased fanouts}
Let $f_0, f_1, ...$ be the sequence of fanouts, where operation $s_i$ is performed on a tree of fanout $f_{i-1}$.  Note that $f_0=B$, that $f_0 \ge f_1 \ge f_2 \ge \cdots$, and   
that the fanout $f_{i - 1}$ must be
determined based only on the first $i - 1$ operations.
When selecting fanouts, we do not need to consider the costs of rebuilds, since in total they
sum to at most $O(L)$. 

We select the fanouts $f_0, f_1, f_2, \ldots$ so that they are always
slightly \emph{query-biased}. In particular, if the first $i$
operations contain $R_i \le L$ queries and $W_i \le N$
inserts/updates, then we select the fanout $f_i$ to be the optimal
fanout for performing $L$ queries and $W_{i}$ inserts/updates. That
is, we always treat the number of queries as $L$, even
if it is much smaller. This rule ensures
that the sequence $f_0 \ge f_1 \ge f_2 \ge \cdots$ is monotone
decreasing.

\paragraph*{Analyzing the performance}
The analysis of each $s_i$ is made slightly
difficult by the fact that $s_i$ is performed with fanout
$f_{i - 1}$ instead of with fanout $f_i$. One useful observation is
that, by slightly tweaking the algorithm, we can achieve performance
asymptotically as good as if each operation $i$ were performed with
fanout $f_i$. This can be enforced by performing the $i$-th operation
with fanout $\sqrt{f_{i - 1}}$ (note that square-rooting the fanout
only hurts the query cost by a constant factor and improves
insert/update cost). As long as
$\sqrt{f_{i - 1}} \le f_i$, then this is asymptotically as good as
using fanout $f_i$. On the other hand, since the sequence
$f_0, f_1, f_2, \ldots$ is monotone decreasing there can only be
$O(\log \log B)$ indices $i$ for which $\sqrt{f_{i - 1}} \ge f_i$ and
the cost of these $O(\log \log B)$ operations is negligible since each
costs $O(\log L)$ I/Os.

Let $f$ be the optimal fanout for performing all of the operations in
$S$ (recall that, without loss of generality, $S$ has at most $L$ queries and $N$ inserts/updates). Each of the fanouts $f_0, f_1, f_2, \ldots$ are
query-biased in the sense that $f_i \ge f$. As a result, we need not
worry about the performance of query operations, that is, we can perform the analysis as though
queries take time $0$. It follows that,
without loss of generality, we may assume that all query operations
are performed at the beginning of the workload (since the positions of
the queries do not affect the fanouts used for the
inserts/updates). Furthermore, we can assume that the number of
queries is precisely $L$, since it turns out that adding $\le L$ queries to a
workload with $N$ inserts/updates does not affect the asymptotic
cost of the workload.

In summary, the following two simplifying assumptions are without loss of generality: that $f_i$ is the fanout used to perform $s_i$, and that $S$ starts with $L$ queries followed only inserts/updates. 
One consequence of the second assumption is that the query-biased rule for selecting the fanouts is equivalent to:
\begin{equation}
  f_i = \operatorname{argmin}_f C_f(\langle s_1, \ldots, s_i \rangle).
  \label{eq:optimal_fi}
\end{equation}
Using \eqref{eq:optimal_fi}, we can prove that
the first $i$ operations cost at most
$C_{f_i}(\langle s_1, \ldots, s_i \rangle)$ (which in turn is $\min_f C_f(\langle s_1,
\ldots, s_i \rangle)$). If we assume that this holds for
$i - 1$, then by induction the cost of the first $i$ operations is at most,
\begin{align*}
  & C_{f_{i - 1}}(\langle s_1, \ldots, s_{i - 1} \rangle) + C_{f_i}(\langle s_i \rangle) \\
   = \ & \min_f \;C_f(\langle s_1, \ldots, s_{i - 1} \rangle) + C_{f_i}(\langle s_i \rangle) \\
   \le \ & C_{f_i} (\langle s_1, \ldots, s_{i - 1} \rangle) + C_{f_i}(\langle s_i \rangle) \\
   = \ & C_{f_i}(\langle s_1, \ldots, s_i \rangle). 
\end{align*}
It follows that the cost of all the operations $s_1, \ldots, s_k$ is
bounded by $\min_f C_f(\langle s_1, \ldots, s_k \rangle)$, as desired.

The guarantee achieved above, in which we are competitive with the best fixed fanout $f$, is the simplest adaptive guarantee
that one could hope for. It does not adapt to the spatial-locality of
where operations are performed in the tree, however,
meaning it is still far from optimal.

\subsection{Considering an OPT with Fixed Pivots and Keys}\label{sec:fixed-pivot}
Before considering dynamic optimality over the class of speed-limited
\bpts, we  consider a simpler class of
adversaries that we call \defn{fixed-pivot \bpts} (or \defn{fixed-pivot trees} for short). A fixed-pivot tree contains some fixed set of $N$ records (where $N$
is assumed to be a power of two) and supports query and update
operations (but not inserts and deletes). A fixed-pivot tree is any
\bpt that satisfies the
\defn{fixed-pivot-structure property:} every internal node $x$ has a
power-of-two fanout, and each of $x$'s children subtrees are exactly
equal-size. The fixed-pivot-structure property ensures that there is essentially
no freedom to select pivots in a fixed-pivot tree. In
particular, each subtree has some power-of-two size $2^j$ and the rank of the subtree's final element (i.e., the pivot for the subtree) is forced to be a multiple of $2^j$. We now describe the \defn{fixed-pivot \jellotree}, which is
$O(\delta^{-1})$-competitive with any
$B^{\consta \delta}$-speed-limited fixed-pivot tree. 

\paragraph*{The structure of a fixed-pivot \jellotree}
One of the challenges of dynamically adapting the fanout of a node $x$
is that, whenever the $x$'s fanout  changes, $x$'s
children must be split or merged accordingly, which consequently
affects their fanouts (and, in particular, when you increase or
decrease the fanout of $x$, the merging/splitting that this action forces
upon the children has the opposite effect of decreasing or increasing
their fanouts, respectively). The interdependence between each node $x$ and its
children complicates the task of dynamically adapting fanouts.

The fixed-pivot \jellotree solves this issue by decomposing the tree
into what we call \defn{supernodes}. Every supernode has a fixed
fanout of $B^{\delta}$ (which we will also assume is a power of
two). Abstractly, each supernode $x$ maintains a buffer of size
$B^{1 + \delta}$, allowing for the supernode to buffer up to $B$
messages for each of its children. This large buffer allows for the
supernode $x$ to be fully insert/update-optimized (meaning that it flushes
messages $B$ at a time) while still having large fanout.

\begin{figure}
    \centering
    \includegraphics{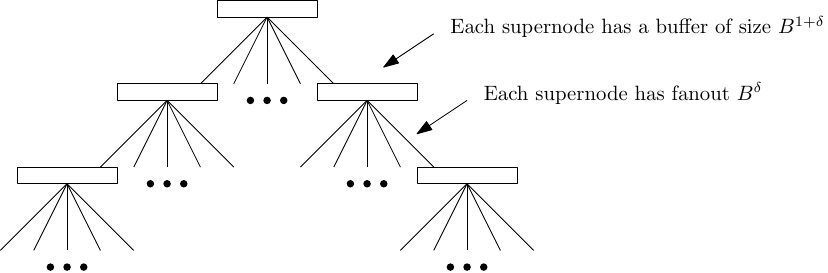}
    \caption{The fixed-pivot \jellotree consists of supernodes with fixed fanouts $B^\delta$. Each supernode has a buffer of size $B^{1 + \delta}$. To handle the fact that the buffer is size $\omega(B)$, each buffer is itself implemented as a fanout-convergent tree.}
    \label{fig:super}
\end{figure}

The downside of a large buffer is that the cost of maintaining and
searching within the buffer is potentially substantial. In order to
optimize this cost, we implement each supernode's buffer as a
fanout-convergent tree (i.e., the data structure from Section \ref{sec:epsilonconv}) that is rebuilt from scratch
every $B^{\delta} \log B^{\delta}$ I/Os.\footnote{We remark that
  supernodes will continue to play a critical role in the design of
  the (non-fixed-pivot) \jellotree later in this overview. The key
  difference will be that, in order to simulate optimal
  pivot-selection, the internal structure of each supernode will
  become substantially more sophisticated.} The supernode structure of a fixed-pivot \jellotree is illustrated in Figure \ref{fig:super}.
  
Whereas the supernode structure of the fixed-pivot \jellotree is
static, the internals of each supernode (and namely the
fanout-convergent tree that implements the buffer) adapt to the
operations that go through the supernode. At first glance, the
fixed-pivot \jellotree may seem quite coarse-grained, in the sense
that each supernode adapts as an entire unit rather than having
individual nodes adapt their fanouts. Nonetheless, we will
see that the adaptive power of the data structure is sufficient to
make it $O(\delta^{-1})$-competitive with any
$B^{\consta \delta}$-speed-limited fixed-pivot tree.

\paragraph*{Imposing a supernode structure on OPT}
Consider a sequence of query and update operations
$S = \langle s_1, s_2, \ldots \rangle$ on the fixed-pivot \jellotree,
and let OPT be the optimal $B^{\consta \delta}$-speed-limited
fixed-pivot tree for $S$.

\begin{figure}
    \centering
    \includegraphics[width=10cm]{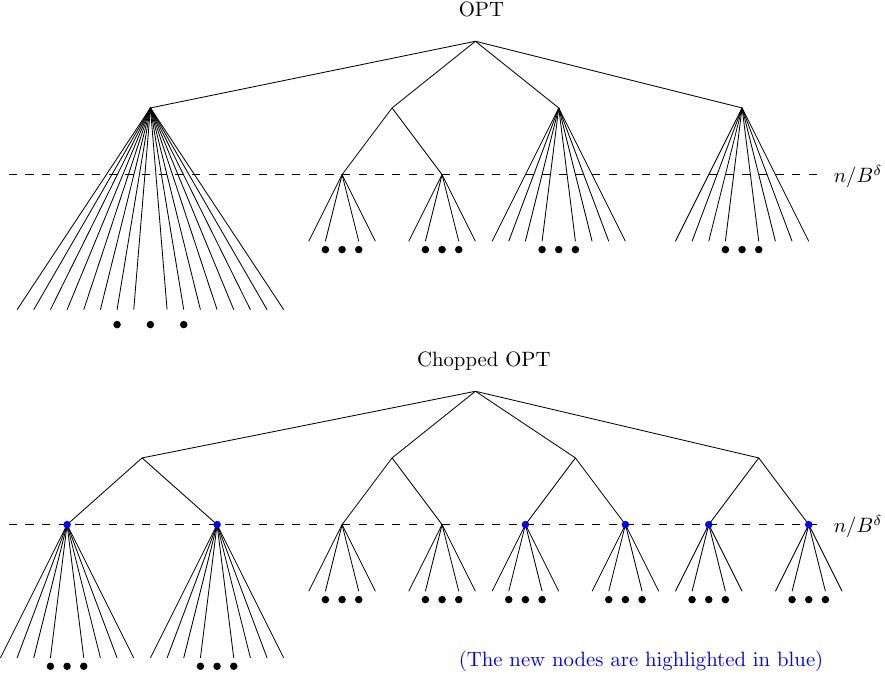}
    \caption{Chopped OPT is constructed from OPT by adding additional nodes so that every root-to-leaf path includes nodes of sizes $N, N / B^\delta, N / B^{2\delta}, \ldots$. Here, we show an example of nodes being added of sizes $N/B^\delta$ (the new nodes are in blue). Notice that, when a node is added, it takes some of the children of its parent.}
    \label{fig:chopped}
\end{figure}

In order to compare the fixed-pivot \jellotree to OPT, we begin by
modifying OPT into a new structure that we call \defn{Chopped OPT} 
that can be partitioned into supernodes. To do this, we add to
OPT a layer of nodes whose sizes (i.e., the number of keys in their
key range) are all exactly $B^{1 + \delta}$, a layer of nodes whose
sizes are all exactly $B^{1 + 2\delta}$, a layer of nodes whose sizes
are all exactly $B^{1 + 3\delta}$, and so on; see Figure \ref{fig:chopped}. (Note that some of these
nodes may already be present in OPT, in which case they need not be
added.)  One
can think of Chopped OPT as consisting of supernodes, where each
supernode has a root of size $B^{1 + h \delta}$ for some $h$ and
leaves of size $B^{1 + (h - 1)\delta}$.

Each root-to-leaf path in Chopped OPT is at most a factor of
$\delta^{-1}$ longer than the same path in OPT. The result is that
Chopped OPT is $O(\delta^{-1})$-competitive with OPT. In order to analyze
the fixed-pivot \jellotree, we show that it is $O(1)$-competitive with
Chopped OPT.

\paragraph*{Competitive analysis against Chopped OPT}
Each supernode in the fixed-pivot \jellotree has a  corresponding supernode 
in Chopped OPT that covers the same key range. 
For each (non-root) supernode $x$ in the fixed-pivot \jellotree,
define the \defn{Chopped-OPT parent} $p(x)$ of $x$ to be the supernode
in Chopped OPT whose key range contains $x$'s key range, and whose
size (in terms of the number of keys in its key range) is
$B^{\delta}$ times larger than $x$'s size. The size requirement means that
$p(x)$ sits one layer higher in Chopped OPT than $x$ sits in the fixed-pivot \jellotree.

In order to analyze the performance of a supernode $x$, there are two
cases to consider, depending on whether Chopped OPT modifies the
structure of $p(x)$ during $x$'s lifetime. We will see that if $p(x)$
is modified then the speed-limitation on Chopped OPT can be used to
amortize the cost incurred by the \jellotree, and otherwise a
competitive analysis can be performed to compare the performance of
supernode $x$ to that of its parent $p(x)$ in Chopped OPT.

The first case is that, at some point during $x$'s life (recall that each supernode $x$ gets rebuilt from scratch after $B^{\delta} \log B^{\delta}$ I/Os), Chopped OPT
modifies $p(x)$. In this case, because Chopped OPT is
$B^{\consta \delta}$-speed-limited, one can think of the modification
of $p(x)$ as costing Chopped OPT $B^{\consta \delta}$ I/Os. On the other hand, the
supernode $p(x)$ in Chopped OPT is a parent supernode for at most
$B^{\delta}$ supernodes $x$ in the \jellotree. Thus we can think of Chopped
OPT as paying $B^{(\consta - 1) \delta}$ I/Os to our supernode $x$. In
other words, the $B^{\consta \delta}$-speed-limitation on Chopped OPT
pays for the $B^{\delta} \log B^{\delta}$ I/Os incurred by $x$ during its life.

The second case is that, over the course of $x$'s lifetime, Chopped
OPT never modifies $p(x)$. In this case, we compare the total cost
incurred by operations in $x$ to the cost incurred by the same
operations in $p(x)$.\footnote{An important subtlety is the effect
  that caching may have on $x$ and $p(x)$. We assume that OPT caches
  all nodes with key-range sizes $N / C$ or larger for some parameter
  $C$, and that the \jellotree caches all nodes with key-range sizes
  $N / (B^{\consta \delta} C)$ or larger. In other words, the
  \jellotree caches $O(1)$ more layers of supernodes than does Chopped
  OPT. The resource augmentation on cache size ensures that, if $x$ is
  (partially) uncached by the \jellotree, then $p(x)$ is (completely)
  uncached by Chopped OPT. In Section \ref{sec:opt}, we also give a
  version of the analysis that does not assume any resource
  augmentation in caching, at the cost of incurring a small additional
  additive cost in the analysis.}
  
Define $S_x$ to be the set of query
and update operations that go through supernode $x$ during $x$'s
lifetime. Whereas the operations in $S_x$ may take different paths than each
other down supernode $x$, all of the operations in $S_x$ take the
\emph{same} root-to-leaf path $P$ through supernode $p(x)$ in Chopped
OPT. We show that the cost incurred by the operations $S_x$ on the
path $P$ in Chopped OPT is minimized by setting all of the fanouts in
$P$ to be equal; we call this the \defn{equal-fanout
  observation}. Note that the equal-fanout observation \emph{does not}
mean that the entire supernode $p(x)$ is optimized by having all of
its fanouts equal; the observation just means that for each
\emph{individual} path in $p(x)$, the cost of the operations that
travel \emph{all the way down} that path would be optimized by setting the
fanouts in that path to be equal (different paths would have different
optimal fanouts, however).

By the equal-fanout observation, the cost that operations $S_x$ incur
in $p(x)$ is asymptotically at least the cost that operations $S_x$ would incur in a
fanout-convergent tree containing $L = \Theta(B^{\delta})$ leaves. On the other hand, the
buffer in $x$ is implemented as a fanout-convergent
tree with $\Theta(B^{\delta})$ leaves. It follows that the cost of operations $S_x$ to
$x$ is $O(1)$-competitive with the cost of operations $S_x$ to $p(x)$.

The analysis described above ignores the fact that update messages may
propagate down the \jellotree at different times than when they
propagate down Chopped OPT. As a result, some of the operations in
$S_x$ may actually remain buffered above supernode $p(x)$ in Chopped
OPT until well after the end of $x$'s lifetime. By the time these
buffered messages make it to $p(x)$, Chopped OPT may have already
modified $p(x)$. It turns out that, whenever this occurs, one can
extend the charging argument from the first case in order to pay for
any I/Os incurred by $x$.

\subsection{The Pivot-Selection Problem}\label{sec:technical-pivot-selection}

\paragraph*{The importance of pivot selection}
The selection of pivots in a \bpt can
have a significant impact on asymptotic performance. Consider, for
example, a sequence of operations
$S = \langle s_1, s_2, \ldots \rangle$, where each operation is
either a query for some key $k_1$ or an update for some key $k_2$,
where $k_2$ is the successor of $k_1$. If the pivots in the
tree are selected independently of the workload $S$, then all of the
operations in $S$ will (most likely) be sent down the same
root-to-leaf path. On the other hand, if the \bpt
uses $k_1$ as a pivot in the root node, then all of the queries to $k_1$
will be sent down one subtree, while all of the
updates to $k_2$ will be sent down another, allowing for the tree to
implement updates in amortized time $O((\log N) / B)$ and queries in amortized time $O(\log_B N)$.
 Therefore, in order to be competitive with OPT, one must be competitive even in the cases were OPT's pivots
 split the workload into natural sub-workloads, each of which is optimized separately with properly selected fanouts. 
As was the
case in this example, the \emph{exact} choice of pivot can be very
important, meaning that there is no room to select a pivot that is
``almost'' in the right position.

\paragraph*{What supernodes must guarantee}
Define the \defn{fanout-convergent cost} for a set of
operations $S$ in a supernode $x$ to be the cost of implementing $S$ in a fanout-convergent tree that has $\Theta(B^{\delta})$ leaves. 
 We say that a key-range
$[k_1, k_2]$ in supernode $x$ \defn{achieves fanout-convergence
over some time window $W$} if the cost of the operations $S$ that
apply to $[k_1, k_2]$ during $W$ is
within a constant factor of the fanout-convergent cost of those
operations.

Consider what goes wrong in the analysis of the
fixed-pivot \jellotree if we allow Chopped OPT to select arbitrary
pivots. Recall that in the competitive analysis, we compare each supernode $x$ to its parent $p(x)$ in
Chopped OPT.

If Chopped OPT is permitted to select arbitrary pivots, however, then
$x$ may actually have \emph{two} parents $p_1(x)$ and $p_2(x)$, each
of which partially overlaps $x$'s key range.\footnote{Because each of
  $x$'s parents covers a larger key-range than $x$,
  $x$ can have at most two parents.} We need the supernode $x$ to achieve
fanout-convergence on \emph{both} of the key ranges
$x \cap p_1(x)$ and $x \cap p_2(x)$ (rather than simply achieving
for the entire key range of $x$).

Since the (non-fixed-pivot) \jellotree does not know what the pivot
$p$ is that separates $p_1(x)$ and $p_2(x)$, the \jellotree must be
able to provide a guarantee for all possible such pivots. 
For any pivot $p$, define the \defn{$p$-split} cost of a supernode $x$ to be the sum of (a) the
  fanout-convergent cost for the operations in $x$ that involve
  keys $\le p$, and (b) the fanout-convergent cost for the
  operations in $x$ that involve keys $> p$. Each supernode $x$ must provide what we call the
\defn{Supernode Guarantee:}
for any $p$, $x$'s total actual cost is $O(1)$-competitive with its $p$-split cost.

\paragraph*{One additional requirement in the supernode guarantee: speed}
One of the aspects of pivot selection that makes it difficult is that
a supernode $x$'s lifetime may be relatively short. In particular,
whenever a supernode $x$'s size changes by a sufficiently large constant factor,
the \jellotree is forced to perform rebalancing on that supernode,
thereby ending $x$'s life. In the worst case, for supernodes $x$ in
the bottom layer of the tree, the lifetime of the supernode may
consist of only $O(B^{1 + \delta})$ inserts (and some potentially small number of
queries), meaning that the total I/O-cost of the supernode could be as
small as $O(B^{\delta} \log B^{\delta})$. Thus convergence to the 
supernode guarantee must be fast.

This issue is further exacerbated by the fact that the supernode
guarantee requires not only pivot selection but also optimal
fanout-convergence on each side of that pivot. But even just the
time to achieve optimal fanout-convergence on a tree with
$B^{\delta}$ leaves, using the approach in Subsection
\ref{sec:epsilonconv}, may take 
$\Theta(B^{\delta} \log B^{\delta})$ I/Os. This means that the natural
approach of achieving fanout convergence from scratch (on both
sides of the pivot) every time that we modify our choice of pivot is
not viable. Instead, the processes of pivot selection and
fanout convergence must interact so that both pieces of the
supernode guarantee can be achieved concurrently within a small
time window.

\paragraph*{The difficulty of a moving target}
One natural approach to pivot-selection is to keep a random sampling
of the operations performed so far and to use this to determine an
approximation of the pivot $p_{\text{opt}}$ that is optimal 
for performing all of the operations so far. If the pivot $p_{\text{opt}}$ is relatively
static over time (e.g., if the operations being performed are drawn
from some fixed stochastic distribution), then such an approach may
work well.  On the other hand, if $p_{\text{opt}}$ shifts over time,
then the approach of ``following'' $p_{\text{opt}}$ fails.

To see why, suppose that $p_{\text{opt}}(t)$ is the optimal pivot
choice for performing the first $t$ operations and that for operation
$t$ we use $p_{\text{opt}}(t - 1)$ as our pivot, i.e., we perfectly follow $p_{\text{opt}}$. Further suppose that
the optimal pivot $p_{\text{opt}}(t)$ places the insert-heavy portion
of the workload on its left side and the query-heavy portion on its
right side. One example of what may happen is that $p_{\text{opt}}$
drifts to the right over time, due to inserts being performed to the
right of where $p_{\text{opt}}$ just was. The result is that, for many
insert operations $t$, the pivot $p_{\text{opt}}(t - 1)$ may be to the
left of the insert-key even though the pivot $p_{\text{opt}}(t)$ is to
the right of the insert-key --- this makes
$p_{\text{opt}}(t - 1)$ a poor pivot to use for operation $t$. One can
attempt to mitigate this by overshooting and using a pivot to the
right of $p_{\text{opt}}(t - 1)$, but this then opens us up to other
vulnerabilities (such as $p_{\text{opt}}$ drifting to the left).

In the next subsection, where we describe our techniques for
implementing the supernode guarantee, we will see an alternative
approach to pivot selection that allows for our performance to
converge to that of the optimal pivot, without having
to ``follow'' it around. We will then also
see how to integrate pivot selection with fanout convergence so
that the supernode guarantee holds even for supernodes
with short lifetimes.

\subsection{Providing the supernode guarantee}\label{sec:technical-supernode}
As is the case for the fixed-pivot \jellotree, the supernodes buffers in the (non-fixed-pivot) \jellotree are implemented with a
tree structure. To avoid ambiguity, we refer to the leaves of this
tree structure as the supernode's \defn{leaves} (even though they are
the children of the supernode, and are therefore other supernodes).

\paragraph*{Simplifying pivot selection by shortcutting leaves} 
A given supernode may have a large number of possible pivots
(especially if the supernode is high in the tree). On the other hand,
as discussed in Subsection \ref{sec:technical-pivot-selection},
picking the wrong pivot (even by just a little) can be disastrous.

We can reduce the effective number of pivot options by adding a 
new mechanism called \defn{shortcutting}. In order to shortcut a leaf $\ell$, we store the
buffer for leaf $\ell$ directly in the root node of the supernode,
meaning that the root takes $1$ more block of space than it would
normally. Whenever a leaf is shortcutted, all messages within the supernode destined for that leaf are stored within the root buffer (and not in any root-to-leaf paths). Queries that go through leaf $\ell$ incur only $O(1)$ I/Os
in the supernode, since they can access $\ell$ directly in the
root. Inserts/updates that go through leaf $\ell$ incur only $O(1/B)$
amortized cost in the supernode, since the leaf gets its own buffer of size $B$ in the root of the supernode.

Because each shortcutted leaf increases the size of the root by $1$
block, we can only support $O(1)$ shortcutted leaves at a time. We prove that, to simulate optimal pivot-selection, one can instead select
$O(1)$ shortcutted leaves in a way so that one of those shortcutted
leaves contains the optimal pivot. This means that, rather than
satisfying the supernode guarantee directly, it suffices to instead
satisfy the following ``shortcutted'' version of the guarantee:
\begin{itemize}
\item \textbf{The Shortcutted Supernode Guarantee: }Consider a
  sequence of operations $S$ on a supernode $x$. For any possible
  shortcutted leaf $\ell$, define the \defn{$\ell$-split cost} of $S$
  to be the sum of (a) the fanout-convergent cost of the
  operations in $S$ that are on keys smaller than those in $\ell$; (b)
  the fanout-convergent cost of the operations in $S$ that are on
  keys larger than those in $\ell$; and (c) the shortcutted cost of
  implementing the operations in $S$ that apply to leaf $\ell$.  The total cost of
  all operations $S$ on a supernode $x$ in its lifetime is guaranteed to be $O(1)$-competitive
  with the $\ell$-split cost of $S$.
\end{itemize}

The shortcutted supernode guarantee implies the standard supernode guarantee,
but the former is more tractable because now, rather than selecting a specific pivot (out of a possibly very large number of options), we only have
to select one of $O(B^\delta)$ leaves to shortcut. Moreover, we get to select \emph{multiple} such 
leaves at a time (we will end up selecting 3 at a time), which will allow for us to perform an algorithm in which we ``chase'' the optimal shortcutted leaf from multiple directions at once.

\paragraph*{An algorithm for shortcut selection} Fix a supernode $x$ and consider the task of implementing the shortcutted supernode guarantee. 
The algorithm breaks the supernode's lifetime into short
\defn{shortcut convergence windows}, where each shortcut
convergence window satisfies the shortcutted supernode guarantee.

\begin{figure}
    \centering
    \includegraphics{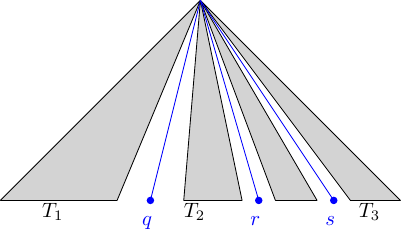}
    \caption{Each supernode selects three children $q, r, s$ (at a time) to shortcut. These children have their buffers stored at the top of the supernode. The four key ranges between $q, r, s$ are implemented as fanout-convergent trees $T_1, T_2, T_3, T_4$.}
    \label{fig:shortcut}
\end{figure}

Each shortcut convergence window is broken into phases, where the
first phase has some length $T$ (in I/Os), and then each subsequent
phase $i$ is defined to consist of $1/8$-th as many I/Os as the sum of
phases $1, 2, \ldots, i - 1$. If we think of I/Os as representing
time, then each phase $i$ extends the length of the shortcut
convergence window by a factor of $1 + 1/8$. At the beginning of each
phase $i$, our algorithm will select three leaves $q < r < s$ to be
shortcutted during that phase. These are the only leaves that are
shortcutted during the phase; see Figure \ref{fig:shortcut}.

At any given time $t$, define the \defn{optimal static shortcut}
$\ell_{\text{opt}}(t)$ for the supernode $x$ to be the leaf $\ell$ that
minimizes the $\ell$-split cost of the operations performed up until time $t$ (during the current shortcut convergence window). During each
shortcut convergence window, we keep track of the optimal static
shortcut $\ell_{\text{opt}}$ as it changes over time.

For now, we will describe the shortcut selection algorithm with two
simplifying assumptions. The first is that the key-ranges
between shortcuts\footnote{By this we mean the four key ranges
  corresponding with the four sets of leaves,
  $\{1, 2, \ldots, q - 1\}$, $\{q + 1, \ldots, r - 1\}$,
  $\{r + 1, \ldots, s - 1\}$, $\{s + 1, \ldots\}$.} each achieve
fanout-convergence during each phase. The second is that
the set of leaves in supernode $x$ does not change during the shortcut convergence window
(i.e., no node-splits occur). Later we will see how to modify the
algorithm to remove both of these assumptions.

We can now describe how the algorithm works. At the beginning of each
phase $i > 1$, there are two \defn{anchor shortcuts} $q$ and $s$ that
have already been shortcutted for all of phase $i - 1$. The key
property that the anchor shortcuts satisfy is that the optimal static
shortcut $\ell_{\text{opt}}$ is between them. The two
anchor shortcuts $q$ and $s$ remain shortcutted for all of phase
$i$. If, at any point during phase $i$, the optimal static shortcut
$\ell_{\text{opt}}$ crosses one of $q$ or $s$ (so that it is no longer
between them), then we \defn{terminate} the entire shortcut
convergence window and begin the next one starting with phase $1$
again --- in a moment, we will argue that whenever the shortcut
convergence window terminates, it satisfies the
shortcutted supernode guarantee.

In addition to shortcutting the anchors $q$ and $s$ during
phase $i$, we also shortcut the leaf $r$ that is half-way between
$q$ and $s$. At the end of the phase, we
then select the anchor shortcuts for phase $i + 1$ to be $\{q, r\}$ if
$\ell_{\text{opt}}$ is between $q$ and $r$, and to be $\{r, s\}$ if
$\ell_{\text{opt}}$ is between $r$ and $s$. The result is that, if the
shortcut convergence window does not terminate during a phase $i$,
then the distance between the anchor shortcuts used in phase $i + 1$
will be half as large as the distance between the anchor shortcuts in
phase $i$.

\paragraph*{Analyzing running time} Before proving the shortcutted
supernode guarantee, we first bound the running time of the shortcut
convergence window $W$. Because the distance between the anchor
shortcuts halves between consecutive phases, the window $W$ is
guaranteed to terminate within $O(\log B^{\delta})$ phases. This means
that the number of I/Os incurred by $x$ is at most
$O(T (1 + 1/8)^{\log B^{\delta}}) \le O(T B^{\delta / 5})$. As long as
$T$ is reasonably small (e.g., less than $B^{\delta / 2}$), then the
length of each shortcut convergence window is small enough to fit in a
supernode's lifetime.

\paragraph*{Proving the shortcutted guarantee} We now argue that, whenever a shortcut convergence window $W$
terminates, the supernode $x$ satisfies the shortcutted
supernode guarantee. Recall that the window 
terminates when the static optimal shortcut $\ell_{\text{opt}}$
crosses over one of the anchors $q$ or $s$ (let's say it crosses $q$). 
The fact that $\ell_{\text{opt}}$ crosses over
$q$ can be used to argue that the $q$-split cost of $x$ during
$W$ is $O(1)$-competitive with the $\ell_{\text{opt}}$-split cost of
the $x$ during $W$. Thus our goal is to compare the total cost
incurred on $x$ during the shortcut convergence window by our data structure
to the $q$-split cost of $x$ for the same time window.

Let $A_1, A_2, \ldots, A_i$ be the costs in I/Os of phases
$1, 2, \ldots, i$ to supernode $x$, and let $B_1, B_2, \ldots, B_i$ be
the $q$-split costs of the operations in each of phases
$1, 2, \ldots, i$ (here $q$ is the shortcut from the final phase
$i$). We wish to show that
$A_1 + \cdots + A_i \le O(B_1 + \cdots + B_i)$. In particular, this
will establish that the cost incurred by $x$ during window $W$ is
constant-competitive with the $q$-split cost during the same window.

Recall that each phase is defined to take a constant-fraction more
I/Os than the previous phase, meaning that $A_1, \ldots, A_i$ are a
geometric series (except for the final term $A_i$ which may be smaller). Thus, rather
then proving that $A_1 + \cdots + A_i \le O(B_1 + \cdots + B_i)$, it
suffices to show that $A_{i - 1} \le O(B_{i - 1})$.

The fact that $q$ is an anchor shortcut in the final phase $i$ implies
that $q$ was shortcutted in $x$ for all of phase $i - 1$. This means that the cost of supernode $x$ to our data structure during phase $i - 1$ is at most the $q$-split cost of the operations in phase $i - 1$, that is, $A_{i - 1} \le O(B_{i - 1})$. As observed above, it follows that $A_1 + \cdots + A_i \le O(B_1 + \cdots + B_i)$, which completes the proof of the
shortcutted supernode guarantee.

\paragraph*{Removing the simplifying assumptions}
At this point, we have completed a high-level overview of how an algorithm can perform shortcut selection in order to achieve the shortcutted supernode guarantee. As noted earlier, however, the analysis makes several significant simplifying assumptions: (1) that the key ranges between shortcuts each achieve fanout-convergence during each phase; and (2) that the set of leaves for supernode $x$ is a static set. Removing these simplifications requires several significant additional technical ideas which we give an overview of in the rest of this subsection.

\paragraph*{Handling a dynamically changing leaf set}
We begin by removing the assumption that $x$'s leaf set is static. For supernodes $x$ in the bottom layer of
the tree (which are the supernodes we will focus on for the rest of
this section), the leaf-set of the supernode may change
dramatically over the course of the supernode's lifetime, due to
inserts causing leaves to split. Thus,
during a given phase $i$ of a shortcut convergence window, the number
of leaves between the two anchor shortcuts $q$ and $s$ may 
increase by more than factor of two. This means that the distance
between the anchor shortcuts that are used in phase $i + 1$ (in terms
of number of leaves between them) could be \emph{larger} than
the distance between the anchor shortcuts in phase $i$. If this
happens repeatedly, then the shortcut convergence window may never
terminate.

To combat this issue, we modify the supernode guarantee as follows.
Rather than comparing the cost of a supernode $x$ to the $p$-split
cost of $x$ for every possible pivot $p$, we only compare the cost of
$x$ to the $p$-split costs for pivots $p$ that were \emph{already
  present in the tree} at the beginning of $x$'s lifetime. We call
these the \defn{valid pivot options} for $x$.

Similarly, we modify the shortcutted supernode guarantee to only
consider the $\ell$-split cost for leaves $\ell$ that contain at least
one valid pivot option. One can show that the weakened version of the two
guarantees still suffices for performing a competitive analysis on the
\jellotree.

In order to provide the new version of the shortcutted supernode
guarantee, we modify the pivot-selection algorithm as follows. Rather
than halving the number of leaves between the anchor shortcuts in each
phase, we instead halve the number of valid pivot options contained in
the leaves between the anchor shortcuts. That is, rather than selecting
the shortcutted leaf $r$ to be half way between the two anchor
shortcuts $q$ and $s$, we instead select $r$ so that it evenly
splits the set of valid pivot options between $q$ and $s$.

With this new algorithm, each shortcut convergence window is
guaranteed to terminate within $O(\log B^{1 + \delta})$ phases. In
order to keep the length of each shortcut convergence window small, we
make it so that each phase $i$ is only an $O(\delta)$-fraction as
large as the sum of phases $1, 2, \ldots, i - 1$ (rather than a
$1/8$-fraction). One side-effect of this is that, for supernodes in
the bottom layer of the tree, the competitive ratio for the supernode
guarantee ends up being $O(\delta^{-1})$ (rather than $O(1)$). Nonetheless, this weakened guarantee still turns out to be sufficient
for the competitive analysis of the \jellotree.\footnote{Importantly,
  it is only in the bottom layer of the tree where we have this
  weakened supernode guarantee. That is, the guarantee continues to
  hold with $O(1)$-competitiveness for all supernodes in higher
  layers.}

\paragraph*{Efficiently combining pivot selection with fanout-convergence}
Next we remove the assumption that, within a given phase of a
shortcut convergence window, the key ranges between consecutive
shortcuts each achieve fanout-convergence. Recall that
a fanout-convergent tree with $B^{\delta}$ leaves requires up to
$\Theta(B^{\delta} \log B^{\delta})$ I/Os to converge. Since we cannot
afford to make the minimum phase length $T$ be
$\Theta(B^{\delta} \log B^{\delta})$, we cannot simply perform
fanout-convergence blindly within each phase.

\begin{figure}
    \centering
    \includegraphics{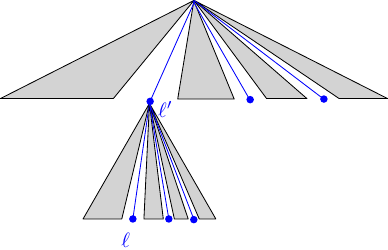}
    \caption{We break each supernode $x$ into two levels, each of which is implemented using micro-supernodes with fanouts $B^{\delta / 2}$. To shortcut a leaf $\ell$ in $x$, we find the micro-supernode $\ell'$ containing $\ell$; we then rebuild the root micro-supernode $r$ so that $\ell'$ is shortcutted in $r$; and finally we rebuild $\ell'$ so that $\ell$ is shortcutted in $\ell'$. Importantly, this process only disrupts fanout-convergence in two of the $\Theta(B^{\delta / 2})$ micro-supernodes.}
    \label{fig:micro}
\end{figure}

In order to perform fanout-convergence and shortcut selection
concurrently, we modify the structure of a supernode as follows. Each
supernode now consists of two layers of \defn{micro-supernodes}, where
each micro-supernode has fanout $\Theta(B^{\delta / 2})$. Each
micro-supernode has the same structure as what we previously gave to
supernodes: each micro-supernode can have up to three shortcut leaves,
and each micro-supernode implements the key ranges between shortcut
leaves as fanout-convergent trees. A leaf $\ell$ in the full supernode $x$ is considered to be shortcutted in $x$ if $\ell$ is shortcutted in the micro-supernode $\ell'$ containing $\ell$, and if $\ell'$ is, in turn, shortcutted in the root micro-supernode; see Figure \ref{fig:micro}.

Whenever the shortcut selection algorithm for a supernode $x$ selects
a new shortcut at the beginning of the phase, it does this by
clobbering and rebuilding only two of the micro-supernodes (specifically, 
the root micro-supernode and one micro-supernode in the bottom layer
of $x$). Critically, this means that the other micro-supernodes
continue to perform fanout convergence without interruption.

If a micro-supernode $y$ survives for
$cB^{\delta / 2} \log B^{\delta / 2}$ I/Os, for some sufficiently large
constant $c$, and the shortcuts of $y$ are never changed by the
shortcut selection algorithm during those I/Os, then each of the
fanout-convergent trees in  $y$
are guaranteed to have achieved fanout convergence (or to have
incurred negligibly few I/Os). In this case, we say that $y$ also achieves fanout convergence.

When a new shortcut is selected, the actual cost of rebuilding the two
micro-supernodes is only $O(B^{\delta /2})$ I/Os. Additionally, the
fact that we clobber two micro-supernodes (possibly before they have a
chance to achieve fanout-convergence) may disrupt fanout convergence for up to
$O(B^{\delta / 2} \log (B^{\delta / 2}))$ I/Os. In this sense, the
total cost of selecting a new shortcut (both the cost in terms of I/Os
expended to rebuild the micro-supernodes, and the cost in terms of the
I/Os that those supernodes had incurred prior to being clobbered) at
the beginning of a phase is $O(B^{\delta / 2} \log (B^{\delta / 2}))$
I/Os. By setting the minimum phase length $T$ to
$cB^{\delta / 2} \log (B^{\delta / 2})$ for a sufficiently large
constant $c$, we can amortize away this cost using the I/Os incurred
in other micro-supernodes during the phase.

\paragraph*{Analyzing pivot selection and fanout-convergence concurrently}
The two-level structure of a supernode, described above, allows for us
to perform shortcut selection and fanout-convergence concurrently
with minimal interference. One issue, however, is that the time frame
in which a given micro-supernode achieves fanout convergence may
overlap multiple phases (and even multiple shortcut convergence
windows) of the shortcut selection algorithm. Thus, the introduction
of micro-supernodes misaligns fanout-convergence and pivot
selection so that the individual shortcut convergence windows
may no longer satisfy the shortcutted supernode guarantee.

In order to get around these issues, we define what we call the
\defn{$p$-re-shortcutted cost} of a supernode $x$ with respect to a
given pivot $p$. Roughly speaking, the re-shortcutted cost of the
supernode $x$ with respect to a pivot $p$ is just the sum of (a) the
actual costs incurred by micro-supernodes in $x$ that do not contain
$p$ in their key range, and (b) the $p$-split cost of each
micro-supernode that does contain $p$ in its key range. Rather than
proving that each shortcut convergence window satisfies the supernode
guarantee, we instead prove a weaker property: for each pivot $p$, the
cost of $x$ in each shortcut convergence window is
$O(1)$-competitive with the $p$-re-shortcutted cost of $x$ during
the same window. Combining this guarantee across all
shortcut convergence windows, we get that the cost of $x$ over its
entire lifetime is $O(1)$-competitive with the $p$-re-shortcutted
cost of $x$ during its entire lifetime. Then, using the fact that
(almost all of) the micro-supernodes in $x$ achieve
fanout-convergence by the end of $x$'s lifetime, we conclude that
the $p$-re-shortcutted cost of $x$ during its lifetime is
constant-competitive with the $p$-split cost of $x$. Thus, even though
each individual shortcut convergence window may not satisfy the
supernode guarantee, the supernode $x$ \emph{does} satisfy the
supernode guarantee over the course of its lifetime.


\begin{thebibliography}{10}

\bibitem{AdelsonLa62}
George~M Adel'son-Vel'skii and Evgenii~Mikhailovich Landis.
\newblock An algorithm for organization of information.
\newblock In {\em Doklady Akademii Nauk}, volume 146, pages 263--266. Russian
  Academy of Sciences, 1962.

\bibitem{AggarwalVi88}
Alok Aggarwal and S.~Vitter, Jeffrey.
\newblock The input/output complexity of sorting and related problems.
\newblock {\em Communications of the ACM}, 31(9):1116--1127, September 1988.

\bibitem{arge1995buffer}
Lars Arge.
\newblock The buffer tree: A new technique for optimal i/o-algorithms.
\newblock In {\em Workshop on Algorithms and Data structures}, pages 334--345.
  Springer, 1995.

\bibitem{AuerCeFrSc95}
Peter Auer, Nicolo Cesa-Bianchi, Yoav Freund, and Robert~E Schapire.
\newblock Gambling in a rigged casino: The adversarial multi-armed bandit
  problem.
\newblock In {\em Proceedings of IEEE 36th Annual Foundations of Computer
  Science}, pages 322--331. IEEE, 1995.

\bibitem{BuadoiuCoDeIa07}
Mihai B{\u{a}}doiu, Richard Cole, Erik~D Demaine, and John Iacono.
\newblock A unified access bound on comparison-based dynamic dictionaries.
\newblock {\em Theoretical Computer Science}, 382(2):86--96, 2007.

\bibitem{Bayer72}
Rudolf Bayer.
\newblock Symmetric binary b-trees: Data structure and maintenance algorithms.
\newblock {\em Acta informatica}, 1(4):290--306, 1972.

\bibitem{BayerMc72}
Rudolf Bayer and Edward~M. McCreight.
\newblock Organization and maintenance of large ordered indexes.
\newblock {\em Acta Informatica}, 1(3):173--189, February 1972.

\bibitem{bender2020flushing}
Michael~A Bender, Rathish Das, Mart{\'\i}n Farach-Colton, Rob Johnson, and
  William Kuszmaul.
\newblock Flushing without cascades.
\newblock In {\em Proceedings of the Fourteenth Annual ACM-SIAM Symposium on
  Discrete Algorithms}, pages 650--669. SIAM, 2020.

\bibitem{BenderFaFi07}
Michael~A. Bender, Martin Farach-Colton, Jeremy~T. Fineman, Yonatan~R. Fogel,
  Bradley~C. Kuszmaul, and Jelani Nelson.
\newblock Cache-oblivious streaming {B}-trees.
\newblock In {\em Proc.\ 19th Annual ACM Symposium on Parallelism in Algorithms
  and Architectures (SPAA)}, pages 81--92, 2007.

\bibitem{BenderFaJaKu15}
Michael~A Bender, Martin Farach-Colton, William Jannen, Rob Johnson, Bradley~C
  Kuszmaul, Donald~E Porter, Jun Yuan, and Yang Zhan.
\newblock And introduction to be-trees and write-optimization.
\newblock {\em Login; Magazine}, 40(5), 2015.

\bibitem{arxiv}
Michael~A Bender, Martin Farach-Colton, and William Kuszmaul.
\newblock What does dynamic optimality mean in external memory?
\newblock {\em arXiv preprint}, 2021.

\bibitem{rocks}
Dhruba Borthakur.
\newblock Under the hood: Building and open-sourcing rocksdb.
\newblock {\em Facebook Engineering Notes}, 2013.

\bibitem{BoseDoIaLa14}
Presenjit Bose, Karim Dou{\"\i}eb, John Iacono, and Stefan Langerman.
\newblock The power and limitations of static binary search trees with lazy
  finger.
\newblock In {\em International Symposium on Algorithms and Computation}, pages
  181--192. Springer, 2014.

\bibitem{BoseDL08}
Prosenjit Bose, Karim Dou{\"\i}eb, and Stefan Langerman.
\newblock Dynamic optimality for skip lists and b-trees.
\newblock In {\em Proceedings of the nineteenth annual ACM-SIAM symposium on
  Discrete algorithms}, pages 1106--1114. Citeseer, 2008.

\bibitem{BrodalFa03c}
Gerth~St{\o}lting Brodal and Rolf Fagerberg.
\newblock Lower bounds for external memory dictionaries.
\newblock In {\em Proc. 14th Annual {ACM-SIAM} Symposium on Discrete Algorithms
  (SODA)}, pages 546--554, 2003.

\bibitem{BuchsbaumGoVeWe00}
Adam~L Buchsbaum, Michael~H Goldwasser, Suresh Venkatasubramanian, and
  Jeffery~R Westbrook.
\newblock On external memory graph traversal.
\newblock In {\em SODA}, pages 859--860, 2000.

\bibitem{ChalermsookGKMS18}
Parinya Chalermsook, Mayank Goswami, L{\'a}szl{\'o} Kozma, Kurt Mehlhorn, and
  Thatchaphol Saranurak.
\newblock Multi-finger binary search trees.
\newblock {\em arXiv preprint arXiv:1809.01759}, 2018.

\bibitem{ColeMiScSi00}
Richard Cole, Bud Mishra, Jeanette Schmidt, and Alan Siegel.
\newblock On the dynamic finger conjecture for splay trees. part i: Splay
  sorting log n-block sequences.
\newblock {\em SIAM Journal on Computing}, 30(1):1--43, 2000.

\bibitem{ColeMiScSi00b}
Richard Cole, Bud Mishra, Jeanette Schmidt, and Alan Siegel.
\newblock On the dynamic finger conjecture for splay trees. part i: Splay
  sorting log n-block sequences.
\newblock {\em SIAM Journal on Computing}, 30(1):1--43, 2000.

\bibitem{Comer79}
Douglas Comer.
\newblock The ubiquitous {B}-tree.
\newblock {\em ACM Computing Surveys}, 11(2):121--137, June 1979.

\bibitem{ConwayFa18}
Alex Conway, Martin Farach-Colton, and Philip Shilane.
\newblock Optimal hashing in external memory.
\newblock In {\em 45th International Colloquium on Automata, Languages, and
  Programming, ICALP 2018}, page~39. Schloss Dagstuhl-Leibniz-Zentrum fur
  Informatik GmbH, Dagstuhl Publishing, 2018.

\bibitem{CormenLeRi01}
T.H. Cormen, C.E. Leiserson, R.L. Rivest, and C.~Stein.
\newblock {\em Introduction To Algorithms}.
\newblock MIT Press, 2001.

\bibitem{DemaineHaIaPa07}
Erik~D Demaine, Dion Harmon, John Iacono, and Mihai P~a ˇ~tra{\c{s}}cu.
\newblock Dynamic optimality—almost.
\newblock {\em SIAM Journal on Computing}, 37(1):240--251, 2007.

\bibitem{DemaineIKL19}
Erik~D Demaine, John Iacono, Grigorios Koumoutsos, and Stefan Langerman.
\newblock Belga b-trees.
\newblock In {\em International Computer Science Symposium in Russia}, pages
  93--105. Springer, 2019.

\bibitem{HowatIaMo13}
John Howat, John Iacono, and Pat Morin.
\newblock The fresh-finger property.
\newblock {\em arXiv preprint arXiv:1302.6914}, 2013.

\bibitem{Iacono01}
John Iacono.
\newblock Alternatives to splay trees with o (log n) worst-case access times.
\newblock In {\em Proceedings of the twelfth annual ACM-SIAM symposium on
  Discrete algorithms}, pages 516--522. Society for Industrial and Applied
  Mathematics, 2001.

\bibitem{Iacono13}
John Iacono.
\newblock In pursuit of the dynamic optimality conjecture.
\newblock In {\em Space-Efficient Data Structures, Streams, and Algorithms},
  pages 236--250. Springer, 2013.

\bibitem{FS2}
William Jannen, Jun Yuan, Yang Zhan, Amogh Akshintala, John Esmet, Yizheng
  Jiao, Ankur Mittal, Prashant Pandey, Phaneendra Reddy, Leif Walsh, et~al.
\newblock Betrfs: Write-optimization in a kernel file system.
\newblock {\em ACM Transactions on Storage (TOS)}, 11(4):1--29, 2015.

\bibitem{ONeilChGa96}
Patrick O'Neil, Edward Cheng, Dieter Gawlic, and Elizabeth O'Neil.
\newblock The log-structured merge-tree ({LSM}-tree).
\newblock {\em Acta Informatica}, 33(4):351--385, 1996.

\bibitem{pebbles}
Pandian Raju, Rohan Kadekodi, Vijay Chidambaram, and Ittai Abraham.
\newblock Pebblesdb: Building key-value stores using fragmented log-structured
  merge trees.
\newblock In {\em Proceedings of the 26th Symposium on Operating Systems
  Principles}, pages 497--514, 2017.

\bibitem{Sherk95}
Murray Sherk.
\newblock Self-adjusting k-ary search trees.
\newblock {\em Journal of Algorithms}, 19(1):25--44, 1995.

\bibitem{SleatorTa83}
Daniel~D Sleator and Robert~Endre Tarjan.
\newblock A data structure for dynamic trees.
\newblock {\em Journal of computer and system sciences}, 26(3):362--391, 1983.

\bibitem{SleatorTa85b}
Daniel~Dominic Sleator and Robert~Endre Tarjan.
\newblock Self-adjusting binary search trees.
\newblock {\em Journal of the ACM (JACM)}, 32(3):652--686, 1985.

\bibitem{Tarjan85}
Robert~Endre Tarjan.
\newblock Sequential access in splay trees takes linear time.
\newblock {\em Combinatorica}, 5(4):367--378, 1985.

\bibitem{Tokutek09}
{Tokutek, Inc.}
\newblock {TokuDB}{\textregistered} for {MySQL} {Storage} {Engine}, 2009.
\newblock \texttt{http://www.tokutek.com}.
\newblock URL: \url{http://www.tokutek.com}.

\bibitem{TokuDB11}
{Tokutek Inc.}
\newblock {TokuDB}.
\newblock \url{http://www.tokutek.com/}, 2011.

\bibitem{FS1}
Jun Yuan, Yang Zhan, William Jannen, Prashant Pandey, Amogh Akshintala, Kanchan
  Chandnani, Pooja Deo, Zardosht Kasheff, Leif Walsh, Michael Bender, et~al.
\newblock Optimizing every operation in a write-optimized file system.
\newblock In {\em 14th $\{$USENIX$\}$ Conference on File and Storage
  Technologies ($\{$FAST$\}$ 16)}, pages 1--14, 2016.

\bibitem{FS6}
Jun Yuan, Yang Zhan, William Jannen, Prashant Pandey, Amogh Akshintala, Kanchan
  Chandnani, Pooja Deo, Zardosht Kasheff, Leif Walsh, Michael~A Bender, et~al.
\newblock Writes wrought right, and other adventures in file system
  optimization.
\newblock {\em ACM Transactions on Storage (TOS)}, 13(1):1--26, 2017.

\bibitem{FS3}
Yang Zhan, Alex Conway, Yizheng Jiao, Eric Knorr, Michael~A Bender, Martin
  Farach-Colton, William Jannen, Rob Johnson, Donald~E Porter, and Jun Yuan.
\newblock The full path to full-path indexing.
\newblock In {\em 16th $\{$USENIX$\}$ Conference on File and Storage
  Technologies ($\{$FAST$\}$ 18)}, pages 123--138, 2018.

\bibitem{FS4}
Yang Zhan, Alexander Conway, Yizheng Jiao, Nirjhar Mukherjee, Ian Groombridge,
  Michael~A Bender, Martin Farach-Colton, William Jannen, Rob Johnson, Donald~E
  Porter, et~al.
\newblock How to copy files.
\newblock In {\em 18th $\{$USENIX$\}$ Conference on File and Storage
  Technologies ($\{$FAST$\}$ 20)}, pages 75--89, 2020.

\bibitem{FS5}
Yang Zhan, Yizheng Jiao, Donald~E Porter, Alex Conway, Eric Knorr, Martin
  Farach-Colton, Michael~A Bender, Jun Yuan, William Jannen, and Rob Johnson.
\newblock Efficient directory mutations in a full-path-indexed file system.
\newblock {\em ACM Transactions on Storage (TOS)}, 14(3):1--27, 2018.

\end{thebibliography}
\end{document}